\documentclass[journal]{IEEEtran}
\IEEEoverridecommandlockouts
\usepackage{cite}
\usepackage{amsmath,amssymb,amsfonts}
\usepackage{soul,balance}
\usepackage{algorithm}
\usepackage{algpseudocode}
\usepackage{graphicx}
\usepackage{textcomp}
\usepackage[dvipsnames]{xcolor}
\usepackage[font=small]{caption}
\usepackage{subcaption}
\usepackage{comment}
\usepackage{bm}
\usepackage{mathtools}
\usepackage{siunitx}
\def\BibTeX{{\rm B\kern-.05em{\sc i\kern-.025em b}\kern-.08em
    T\kern-.1667em\lower.7ex\hbox{E}\kern-.125emX}}

\begin{document}

\title{EMF-Compliant Power Control in Cell-Free Massive MIMO: Model-Based \\ and Data-Driven Approaches}
\author{Sergi~Liesegang,~\IEEEmembership{Member,~IEEE,} and Stefano~Buzzi,~\IEEEmembership{Senior Member,~IEEE}
    \thanks{A preliminary and compressed version of this paper has been presented in the conference paper \cite{Lie24}.}
    \thanks{The European Union (EU) supported the work of S. Liesegang under the MSCA Postdoctoral Fellowship DIRACFEC (grant agreement No. 101108043). The EU also supported the work of S. Buzzi under the Italian National Recovery and Resilience Plan (NRRP) of NextGenerationEU, partnership on “Telecommunications of the Future” (PE00000001 - program “RESTART”, Structural Project 6GWINET, Cascade Call SPARKS). Views and opinions expressed are those of the authors and do not necessarily reflect those of the EU. The EU cannot be held responsible for them.}
    \thanks{
    The authors are with the Department	of Electrical and Information Engineering (DIEI), University of Cassino and Southern Latium (UNICAS), 03043 Cassino, Italy, and the \textit{Consorzio Nazionale Interuniversitario per le Telecomunicazioni} (CNIT), 43124 Parma, Italy. S. Buzzi is also affiliated with the Department of Electronics, Information, and Bioengineering (DEIB), Politecnico di Milano (PoliMI), 20122 Milano, Italy.}  
}

\maketitle

\begin{abstract}
The impressive growth of wireless data networks has recently led to increased attention to the issue of electromagnetic pollution and the fulfillment of electromagnetic field (EMF) exposure limits. This paper tackles the problem of power control in user-centric cell-free massive multiple-input-multiple-output (CF-mMIMO) systems under EMF constraints. Specifically, the power allocation maximizing the minimum data rate across users is derived for both the uplink and the downlink. To solve such optimization problems, two approaches are proposed, i.e., model-based and data-driven. The proposed model-based solutions for the downlink utilize successive convex optimization and the log-sum-exp approximation for the minimum of a discrete set, whereas ordinary techniques are employed for the uplink. With regard to data-driven solutions,  solutions based on both end-to-end architectures and deep unfolding techniques are explored. Extensive numerical results confirm that the proposed model-based solutions effectively fulfill the EMF constraints while ensuring very good performance; moreover, the results show that the proposed data-driven approaches can tightly approximate the performance of model-based solutions but with much lower computational complexity.  
\end{abstract}


\section{Introduction}
In light of the remarkable proliferation of wireless communications, the global apprehension regarding health implications associated with electromagnetic field (EMF) exposure is on the rise \cite{Zap22}. The surging demand for broadband services necessitates denser deployments and higher frequencies \cite{Tat21}, compelling the incorporation of EMF exposure constraints in the planning of wireless networks \cite{Pat23}. The latest research from the International Commission on Non-Ionizing Radiation Protection (ICNIRP) and the United States Federal Communications Commission (FCC) has outlined various criteria to limit user radiation and mitigate health concerns (cf. \cite{Chi21}). 

Typically, user radiation is quantified using metrics such as specific absorption rate (SAR) and incident power density (IPD) \cite{ITU20}. These capture the characteristics of the propagation environment and quantify EMF exposure both (i) on the human body (or specific body parts) in W/kg and (ii) across a defined coverage area in W/m\textsuperscript{2}. For short distances (less than 20 cm) and low frequencies (below 6 GHz), the skin penetration is deeper, making SAR the dominant metric for uplink (UL) radiation exposure \cite{Pso22}. Conversely, IPD restrictions are commonly applied in the downlink (DL) transmissions, where EMF absorption is primarily superficial \cite{ICNIRP20}.

An essential aspect of 5G wireless technologies involves the deployment of base stations equipped with numerous transmitting antennas, known as massive multiple-input-multiple-output (mMIMO) \cite{Lu14}. In traditional cellular systems, edge users contend with both inter-cell interference and poor channel conditions, resulting in performance degradation \cite{Buz17}. To address the issues faced by these cell-edge users, an alternative solution, distributing antennas through access points (APs) and enabling a user-centric design of the network, has been deeply investigated over the last decade, leading to the concept of \textit{cell-free} mMIMO (CF-mMIMO) \cite{Int19}. CF-mMIMO eliminates cell borders, ensuring a high quality of service (QoS) for all users, and emerges as a prominent technology for future 6G services. Moreover, as shown in our preliminary work \cite{Lie24}, it permits fulfilling users' QoS requirements with much lower EMF than traditional multi-cell mMIMO networks. 

One challenge for the CF-mMIMO network is the fact that, owing to the inherent nature of the signal model, the formulation of QoS power control policies is usually nonconvex. This intricacy makes the problem solely tractable via iterative procedures like successive convex optimization (SCO) \cite{Lie23}. Regrettably, these methodologies entail an immense computational complexity that rapidly escalates with the dimension of the networks, a fact that poses the question of their feasibility for real-time implementations \cite{Hoy21}. 

To overcome this challenge, we use machine learning (ML) techniques for the network design and reformulate the problem using (fully-connected) deep neural networks (DNNs) \cite{Guo22}. Accordingly, we employ model-based solutions to generate a comprehensive database for the DNNs. Two different strategies are contemplated: conventional end-to-end training \cite{Dah21} and novel deep unfolding (or algorithm unrolling) \cite{Jag21}. 

Unfortunately, SCO schemes relying on feasibility problems (which normally arise in maximin problems) are not compatible with algorithm unrolling because they are ill-defined in DNN architectures \cite{Pel22}. The standard QoS formulation is then transformed using the log-sum-exp (LSE) approximation for the optimization to be unfoldable. The accuracy of this approach is experimentally assessed via simulations. Finally, all the ML supervised methods undergo a comparative analysis against their theoretical counterparts, evaluating their performance in terms of QoS, EMF, and computational complexity. 

\subsection{State of the Art}
The inclusion of EMF constraints into the design of wireless networks has been around for many years \cite{Hoc14}. Owing to its growing importance among the worldwide public, numerous relevant works can be found in the context of cellular networks. For instance, the authors of \cite{Cas21} explore algorithms for maximizing the rate over time in multi-antenna single-user systems while staying within SAR limits. Based on perfect and casual channel state information (CSI), the performance of optimal, heuristic, and asymptotic methods is analyzed with respect to (w.r.t.) different regimes of the so-called SAR-to-noise ratio. This newly introduced metric can be useful to characterize the tightness of the SAR constraints. The UL is also addressed in \cite{Jia22}, where the authors concentrate on the spectral efficiency (SE) problem in hybrid RIS (reconfigurable intelligent surfaces) and DMA (dynamic metasurface antennas) assisted multi-user MIMO systems. Considering full and partial CSI assumptions, the transmit covariance, RIS phase shifts, and DMA weights are jointly optimized. In \cite{Gon24}, the focus is shifted towards the DL, where the authors derive performance metrics to jointly quantify the EMF exposure and coverage in urban deployments using Ginibre and Poisson point processes. The authors also validate their findings numerically with realistic datasets. A much broader survey is available in \cite{Fay23}.

However, except for our early study in \cite{Lie24}, where we indeed compare the distributed and centralized mMIMO architectures, little to no other investigations have been conducted in the framework of CF-mMIMO systems. Only the authors of \cite{Wia23} have analyzed the DL throughput and IPD with the help of stochastic geometry tools. The paper computes the first statistical moments, along with the marginal and joint distributions, which are used to reveal the achievable trade-offs between these quantities under maximum ratio transmission.  

On the other hand, ML for wireless communications has also drawn substantial interest in the last years (cf. \cite{Bjo20}). In \cite{Zah23}, the authors examine the DL power allocation in CF-mMIMO systems and formulate the sum-SE and proportional fairness maximizations. The problems are first solved using the WMMSE (weighted minimum mean square error)-ADMM (alternating direction method of multipliers) algorithm to generate the training data. Distributed and clustered DNNs are then proposed to approximate the power coefficients using only local statistical information. The paper \cite{Tua22} presents a distributed learning-based framework using graph neural networks to maximize the sum ergodic rate in CF. To avoid heavy computations, the proposed approach allocates transmit power using locally trained models. The authors of \cite{Men24} explore a deep reinforcement learning (DRL) approach for UL power control in CF-mMIMO. The slow convergence of traditional DRL methods is overcome by means of a prioritized sampling technique in dynamic environments with user mobility and device activity changes. The UL in CF is also investigated in \cite{Usm24}, where the goal is to jointly optimize the pilot assignment and power control with unsupervised learning. In particular, the authors design a multi-task DNN with a custom loss function to maximize the minimum rate subject to a power budget. For a more elaborate review, please refer to \cite{Gko23}.

Recently, deep unfolding has gained popularity among the research community \cite{Mon21}. As an example, the paper \cite{Shl21} introduces a data-driven multi-user MIMO receiver that performs joint symbol detection under CSI uncertainties. The authors integrate ML into the iterative soft interference cancellation algorithm to handle both linear and non-linear channels. In \cite{Li23}, a graph-based trainable framework for maximizing weighted sum energy efficiency is implemented. The authors unfold the successive concave approximation for the power allocation via graph convolutional neural networks and demonstrate its strong generalization to diverse topologies. The authors of \cite{Ngu24} propose a fast and efficient hybrid beamforming design for joint communications and sensing in mMIMO systems. They introduce a modified projected gradient ascent method and unfold it into a trainable deep learning framework. A detailed overview of other works related to algorithm unrolling is provided in \cite{Dek25}.

Despite notable progress, however, the last referenced papers focus on cellular networks. To the best of our knowledge, no studies have been reported so far in the direction of power control for CF-mMIMO systems with both QoS and EMF requirements.

\subsection{Contributions} \label{sec:1.2}
The purpose of this paper, which can be cast as an extension of the preliminary results presented by the same authors in the conference paper \cite{Lie24}, is to fill the previous gaps in the existing literature. More precisely, the main contributions of this work can be listed in the following:
\begin{enumerate}
    \item We introduce a system model to characterize the data transmission and user radiation occurring in DL and UL scenarios of a user-centric CF-mMIMO system. IPD and SAR are considered to measure the perceived EMF exposure in the DL and UL, respectively.
    \item We present a power control framework for maximizing QoS rate under EMF constraints. In the DL case, SCO is used to find feasible power coefficients.
    \item We implement a data-driven approach relying on fully-connected DNNs, trained in an end-to-end fashion using the previous model-based solutions in the DL and UL. Heuristic power control mechanisms are employed as inputs to streamline the training, and robust normalization is applied to cope with the presence of outliers.
    \item We derive an unfoldable DL allocation through the LSE approximation and SCO techniques. The outputs at each iteration serve as databases for algorithm unrolling.    
    \item We propose a deep unfolded scheme that is built upon a concatenated series of simple DNNs, which are then sequentially trained to alleviate the computational load. The structure of each DNN is adjusted to resemble an iteration of the unfolded policy in the DL setup.
\end{enumerate}

The study is completed with extensive numerical experiments, which evaluate the effectiveness of the proposed design in terms of user data rate, EMF radiation, and computational complexity. The simulations demonstrate the satisfactory performance of the proposed ML algorithms. These outcomes also allow us to unveil the crucial role that expert knowledge plays in the learning process and to reveal existing trade-offs between communication and exposure in various scenarios.

\subsection{Organization} \label{sec:1.3}
This paper is structured as follows. Section~\ref{sec:2} introduces the system models and EMF exposures for DL and UL. Section~\ref{sec:3} formulates the QoS optimization problems, while their solution is derived in Section~\ref{sec:4}. Sections~\ref{sec:5} and \ref{sec:6} present the data-driven implementations, namely end-to-end training and deep unfolding, respectively. Section~\ref{sec:7} is devoted to the numerical experiments. Section~\ref{sec:8} concludes the work.

\subsection{Notation} \label{sec:1.4}
In this work, scalars, vectors, and matrices are denoted by italic, boldface lower-case, and upper-case letters, respectively. $\mathbf{0}_m$ denotes the all-zeros vector of length $m$, $\mathbf{I}_m$ denotes the identity matrix of size $m \times m$, and $\mathbb{C}^{m \times n}$ denotes the $m$ by $n$ dimensional complex space. The transpose, Hermitian, inverse, trace, and expectation operators are denoted by $(\cdot)^{\rm T}$, $(\cdot)^{\rm H}$, $(\cdot)^{-1}$, $\textrm{tr}(\cdot)$, and $\mathbb{E}[\cdot]$ respectively. $\mathcal{CN}(\cdot,\cdot)$ denotes the complex proper Gaussian distribution.

\section{System Model}\label{sec:2}
\begin{figure}[t]
\centerline{\includegraphics[trim={0 0 0cm 0cm},clip=true, scale=0.25]{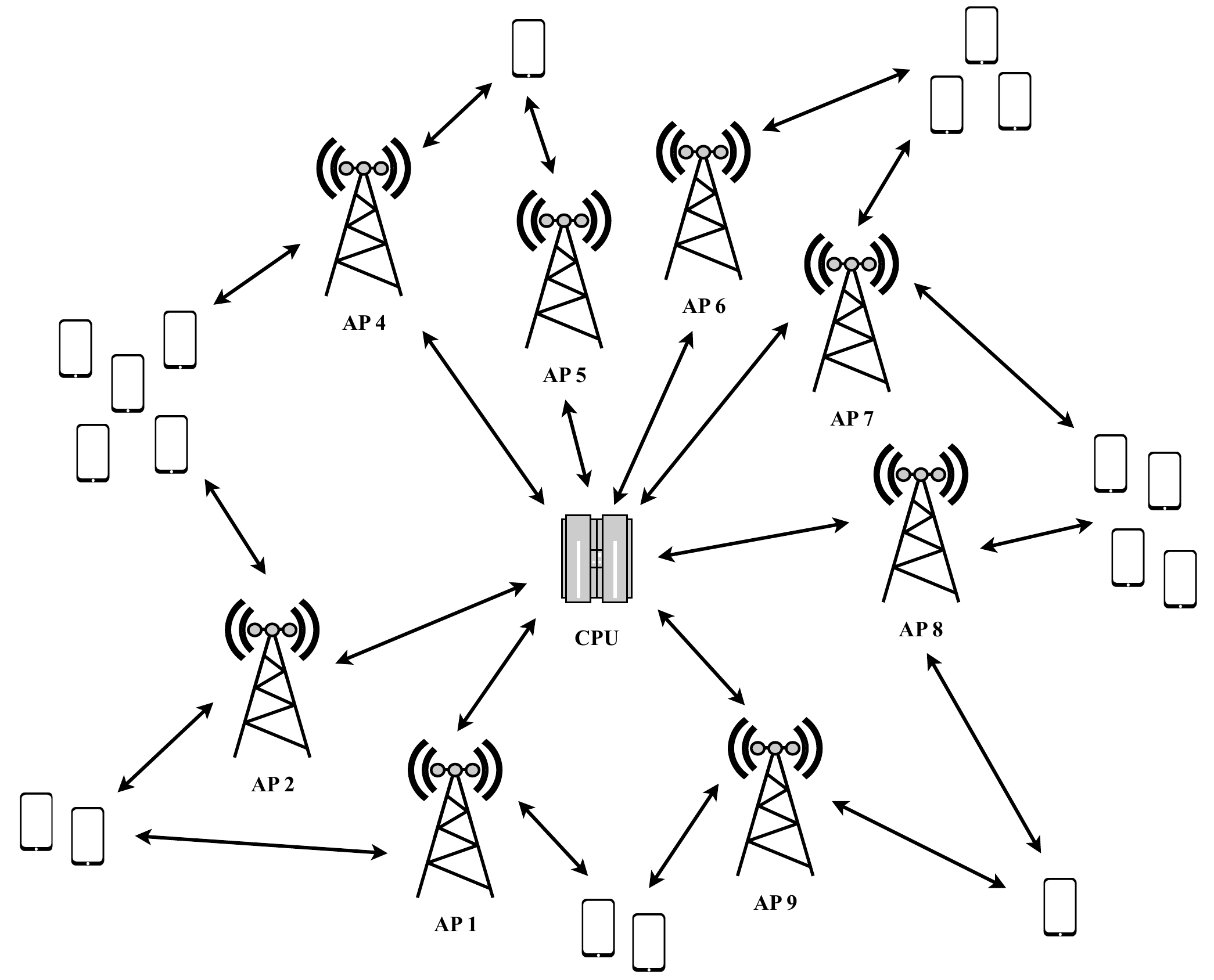}}
\caption{CF-mMIMO deployment where $K = 18$ UEs are connected to $N = 2$ APs, out of a total of $M = 9$, each with $L = 3$ antennas.}
\label{fig:1}
\end{figure}

We adopt a scenario analogous to the one outlined in \cite{Lie24}, focusing on a cell-free deployment comprising $K$ single-antenna user equipments (UEs) being served by $M$ APs equipped with $L$ antennas and connected to a central processing unit (CPU) via fronthaul links with unlimited capacity\footnote{As in \cite{Lie24}, the focus is on the impact of radiation constraints; the extension to a capacity-limited fronthaul is out of the scope of this study.}. As elaborated later, each UE is linked with only a subset of $N$ APs, e.g., those with the highest large-scale fading (LSF) coefficients \cite{Elw23}. An illustrative example is pictured in Fig.~\ref{fig:1}.  

\subsection{Downlink Transmission}\label{sec:2.1}
Under a non-orthogonal communication, the discrete-time signal transmitted from AP $m$ can be expressed as per \cite{Dem21}
\begin{equation}
    \mathbf{x}_m = \sum_{k = 1}^K a_{k,m} \sqrt{p_{k,m}} \mathbf{b}_{k,m}s_k,
    \label{eq:1}
\end{equation}
where $a_{k,m} = 1$ indicates UE $k$ is connected to AP $m$ and $a_{k,m} = 0$ otherwise. The set $\{p_{k,m}\}$ denotes the DL power control coefficients, and the unit vector $\mathbf{b}_{k,m} \in \mathbb{C}^{L}$ represents the beamforming scheme. We also assume standard complex Gaussian transmit signals, i.e., $s_k \sim \mathcal{CN}(0,1)$.

The signal received at UE $k$ is written as
\begin{equation}
    y_k = \sum_{m = 1}^M \mathbf{h}_{k,m}^{\rm H}\mathbf{x}_m + w_k,
    \label{eq:2}
\end{equation}
where $\mathbf{h}_{k,m} \in \mathbb{C}^{L}$ is the UL channel to AP $m$ and $w_k$ is the thermal noise with variance $\sigma_k^2$, i.e., $w_k \sim \mathcal{CN}(0,\sigma_k^2)$. Note that, in line with CF-mMIMO literature (cf. \cite{Buz17}), here we have assumed channel reciprocity thanks to the use of time-division duplexing (TDD). 

As a result, assuming independent transmit signals $s_k$, the signal-to-interference-plus-noise ratio (SINR) at the desired UE can be expressed as follows:
\begin{equation}
    \gamma_k = \frac{\displaystyle \left\vert \sum_{m = 1}^M a_{k,m} \sqrt{p_{k,m}} \mathbf{h}_{k,m}^{\rm H} \mathbf{b}_{k,m} \right\vert^2}{\displaystyle \sum_{j \neq k}\left\vert \sum_{m = 1}^M a_{j,m} \sqrt{p_{j,m}} \mathbf{h}_{k,m}^{\rm H} \mathbf{b}_{j,m} \right\vert^2 + \sigma_k^2}.
    \label{eq:3}
\end{equation}

Note that the instantaneous SINR reported above holds under the assumption of perfect local CSI \cite{Int19}. To simplify the analysis and provide an initial perspective on CF-mMIMO performance under EMF constraints, we begin by adopting this ideal assumption. Estimation errors are subsequently incorporated in the numerical simulations.

When neglecting the noise radiation, the IPD can be modeled as the power components from intended and interfering signals perceived at the UE's position \cite{Pso22, Gon24}:
\begin{equation}
    \xi_k = \frac{4 \pi}{\lambda^2} \sum_{j = 1}^K \left\vert \sum_{m = 1}^M a_{j,m} \sqrt{p_{j,m}} \mathbf{h}_{k,m}^{\rm H} \mathbf{b}_{j,m} \right\vert^2,
    \label{eq:4}
\end{equation}
with $\lambda$ the carrier wavelength.

\subsection{Uplink Transmission}\label{sec:2.2}
The signal received at AP $m$ is given by \cite{Elw23}
\begin{equation}
    \mathbf{r}_m = \sum_{k = 1}^K \mathbf{h}_{k,m}z_k + \mathbf{n}_m,
    \label{eq:5}
\end{equation}
where $z_k$ is the UE transmit signal and $\mathbf{n}_m$ is the additive white Gaussian noise (AWGN), i.e., $\mathbf{n}_m \sim \mathcal{CN}(\mathbf{0}_{L},\eta_m^2 \mathbf{I}_{L})$. Again, we consider $z_k$ follows a complex normal distribution with zero mean and power $q_k$, i.e., $z_k \sim \mathcal{CN}(0,q_k)$.

Using linear filters $\mathbf{f}_{k,m} \in \mathbb{C}^{L}$, APs detect the messages from their associated UEs and send them towards the CPU for jointly decoding signal $z_k$. Under an aggregate estimation $\hat{z}_k = \sum_{m=1}^M a_{k,m} \mathbf{f}_{k,m}^{\rm H} \mathbf{r}_m$, the SINR results \cite{Dem21}
\begin{equation}
    \rho_k = \frac{\displaystyle q_k \left\vert \sum_{m = 1}^M a_{k,m} \mathbf{f}_{k,m}^{\rm H}  \mathbf{h}_{k,m} \right\vert^2}{\displaystyle \sum_{j \neq k}q_j\left\vert \sum_{m = 1}^M a_{k,m} \mathbf{f}_{k,m}^{\rm H} \mathbf{h}_{j,m} \right\vert^2 + \sum_{m = 1}^M a_{k,m} \eta_m^2 \| \mathbf{f}_{k,m} \|_2^2 }.
    \label{eq:6}
\end{equation}

Lastly, the EMF exposure reads as \cite{Zap22, Cas21}
\begin{equation}
    \varepsilon_{k,n} = b_{k,n} q_k,
    \label{eq:7}
\end{equation}
where $b_{k,n}$ are the SAR coefficients associated with all body parts (e.g., head, chest, etc.). 

\section{Problem Formulation}\label{sec:3}
This study aims to develop allocation strategies that maximize the minimum data rate in both scenarios, DL and UL, thereby ensuring a certain QoS for all UEs while adhering to EMF constraints. The optimization can be formulated as
\begin{equation}
    \max_{\mathcal{P}} \, \min_k \, \, R_k\left(\mathcal{P}\right) \quad \textrm{s.t.} \quad \mathcal{C}\left(\mathcal{P}\right),
    \label{eq:8}
\end{equation}
with $\mathcal{P}$ the set of power coefficients, $R_k(\mathcal{P})$ the UE throughput, and $\mathcal{C}(\mathcal{P})$ the constraint set. As previously mentioned, the above problem will be first addressed through model-based solutions and later through data-driven approaches. This is yet discussed in Sections~\ref{sec:4} and \ref{sec:5}, respectively.

\subsection{Downlink Scenario}\label{sec:3.1}
Here, the AP's power coefficients $\mathcal{P} = \{p_{k,m}\}$ are designed to optimize the UE data rate
\begin{equation}
    R_k\left(\mathcal{P}\right) = \frac{\tau_d}{\tau_c} B \log_2 \left(1 + \gamma_k\left(\{p_{k,m}\}\right)\right),
    \label{eq:9}
\end{equation}
subject to the following set of constraints $\mathcal{C}(\mathcal{P})$
\begin{equation}
    \begin{aligned}
        C1:& \quad p_{k,m} \geq 0, \quad \forall k,m \\
        C2:& \quad \sum_{k = 1}^K a_{k,m} p_{k,m} \| \mathbf{b}_{k,m} \|_2^2 \leq P_m, \quad \forall m \\
        C3:& \quad \xi_k \leq I_k, \quad \forall k,
    \end{aligned}
    \label{eq:10}
\end{equation}
where $B$ is the system's bandwidth and $\tau_d$ is the part of the coherence time $\tau_c$ dedicated to DL transmission. Following the TDD protocol, we will have $\tau_d + \tau_u \leq \tau_c$, with $\tau_u$ the portion used for UL communication \cite{Elw23}. Besides, note that $C1$ and $C2$ represent the power limits, while $C3$ constrains the EMF exposure to a maximum bound.

\subsection{Uplink Scenario}\label{sec:3.2}
In this scenario, since the power set is directly $\mathcal{P} = \{q_k\}$, the UE throughput will be defined by
\begin{equation}
    R_k\left(\mathcal{P}\right) = \frac{\tau_u}{\tau_c} B\log_2 \left(1 + \rho_k\left(\{q_k\}\right)\right),
    \label{eq:11}
\end{equation}
and, hence, the constraint set $\mathcal{C}(\mathcal{P})$ will be
\begin{equation}
    \begin{aligned} 
        C1:& \quad 0 \leq q_k \leq Q_k, \quad \forall k \\
        C2:& \quad \varepsilon_{k,n} \leq E_{k,n}, \quad \forall k,n,
    \end{aligned}
    \label{eq:12}
\end{equation}
with $C1$ and $C2$ limiting the power budget and the maximum perceived radiation, respectively. For a thorough analysis, a more detailed explanation is not elaborated until Section~\ref{sec:6}, where all the involved parameters are explicitly delineated.

\section{Model-Based Solution}\label{sec:4}
The optimization problem's convexity varies depending on the type of data transmission, either DL or UL. The following subsections differentiate between these configurations.

Due to the nature of \eqref{eq:8}, suboptimal methods are often necessary to find a feasible solution in the DL. In the sequel, we introduce an iterative procedure based on SCO that converges to a stationary point \cite{Lie23}. Concisely, at the $i$-th iteration, the set of power coefficients $\mathcal{P} \equiv \mathcal{P}^{(i)}$ is gradually updated from the previous point $\mathcal{P}^{(i -1)}$ until convergence is accomplished. For that endeavor, we first reformulate the original problem by moving the objective function into a new constraint (i.e., the standard epigraph form), and then approximate the resulting set $\mathcal{C}(\mathcal{P})$ with appropriate convex functions. 

In contrast, as we will see next, the resulting problem in the UL setup is convex. Thus, a globally optimal solution can be achieved using standard optimization techniques \cite{Ngo17}.

\subsection{Downlink Power Control}\label{sec:4.1}
Let $d_{k,m} \triangleq \sqrt{p_{k,m}} \geq 0$ be the new design variable. This way, one can be show problem \eqref{eq:8} is equivalent to (cf. \cite{Boy04})

\noindent
\begin{equation}
    \begin{aligned}
        \max_{\{d_{k,m}\},\delta} \, \, & \delta \\
        \textrm{s.t.} \quad & C1: d_{k,m} \geq 0, \quad \forall k,m\\ 
        &C2: \sum_{k = 1}^K a_{k,m} d_{k,m}^2 \leq P_m, \quad \forall m \\ 
        &C3: \frac{4 \pi}{\lambda^2} \sum_{j = 1}^K \left\vert \sum_{m = 1}^M a_{j,m} d_{j,m} \mathbf{h}_{k,m}^{\rm H} \mathbf{b}_{j,m} \right\vert^2 \leq I_k, \quad \forall k \\
        &C4: \displaystyle\frac{\tau_d}{\tau_c} B \log_2 \left(1 + \gamma_k\left(\{d_{k,m}\}\right) \right) \geq \delta, \quad \forall k,
    \end{aligned}
    \label{eq:13}    
\end{equation}
where we introduced the auxiliary variable $\delta$ to convert the non-concave objective function into the new constraint $C4$, which remains nonconvex. Given this, we proceed as follows.

Thanks to the logarithm being a monotonically increasing function, $C4$ can be written subsequently:
\begin{equation}
    \begin{aligned}
        \delta \underbrace{ \left(\sum_{j \neq k}\left\vert \sum_{m = 1}^M a_{j,m} d_{j,m} \mathbf{h}_{k,m}^{\rm H} \mathbf{b}_{j,m} \right\vert^2 + \sigma_k^2\right)}_{\triangleq f_k\left(\left[\mathbf{d}_1,\ldots,\mathbf{d}_{k - 1},\mathbf{d}_{k + 1},\ldots,\mathbf{d}_K\right]\right)}& \\
        - \underbrace{\left\vert \sum_{m = 1}^M a_{k,m} d_{k,m} \mathbf{h}_{k,m}^{\rm H} \mathbf{b}_{k,m} \right\vert^2}_{\triangleq g_k\left(\mathbf{d}_k\right)}& \leq 0, \quad \forall k,
    \end{aligned}
    \label{eq:14}
\end{equation}
i.e., a difference of convex functions w.r.t. the “amplitude” vectors $\mathbf{d}_k \triangleq [d_k,\ldots,d_{k, M}]^{\rm T}$.

Problems involving nonconvex constraints lack analytical closed-form solutions. However, using SCO, we can achieve a local optimum. In a nutshell, for a fixed $\delta$, the optimization is decomposed into a sequence of subproblems addressed iteratively. Each must be globally solved to guarantee convergence, meaning the second term $g_k(\mathbf{d}_k)$ in $C4$ must be approximated by a surrogate function \cite{Sun17}. Finally, a bisection search can be used to find the optimal value of $\delta$ \cite{Ngo17}.

Among others, a popular strategy is to linearize the function $g_k(\mathbf{d}_k)$ to convexify the constraint in \eqref{eq:14}. Applying the first-order Taylor expansions at the previous feasible point, i.e., $\mathbf{d}_k^{(i - 1)}$, we obtain the following lower bound:
\begin{equation}
    \begin{aligned}
    g_k\left(\mathbf{d}_k\right) &\geq g_k\left(\mathbf{d}_k^{(i - 1)}\right) + \nabla g_k\left(\mathbf{d}_k^{(i - 1)}\right)^{\rm T}\left(\mathbf{d}_k- \mathbf{d}_k^{(i - 1)}\right) \\
    &\triangleq \tilde{g}_k\left(\mathbf{d}_k,\mathbf{d}_k^{(i - 1)}\right),
    \end{aligned}
    \label{eq:15}
\end{equation}
where the gradient reads as $\nabla g_k(\mathbf{d}_k) = 2 \textrm{Re}\{\mathbf{c}_k\mathbf{c}_k^{\rm H}\}\mathbf{d}_k$, with $\mathbf{c}_k \triangleq [a_k\mathbf{h}_{k,1}^{\rm H} \mathbf{b}_{k,1},\ldots, a_{k, M}\mathbf{h}_{k, M}^{\rm H} \mathbf{b}_{k, M}]^{\rm T}$. 

By defining $\bar{\mathbf{d}}_k \triangleq [\mathbf{d}_1,\ldots,\mathbf{d}_{k - 1},\mathbf{d}_{k + 1},\ldots,\mathbf{d}_K]$, at the $i$-th iteration we will tackle the feasibility problem below \cite{Boy04}:
\begin{equation}
\begin{aligned}
        \textrm{find} \, \, & \{d_{k,m}\} \quad \textrm{s.t.} \quad C1,C2,C3 \\
        & C4: \delta f_k\left(\bar{\mathbf{d}}_k\right) - \tilde{g}_k\left(\mathbf{d}_k,\mathbf{d}_k^{(i - 1)}\right) \leq 0, \quad \forall k.
    \end{aligned}
    \label{eq:16}      
\end{equation}

This way, given a $\delta$ determined by bisection, we end up with a series of subproblems that are worst-case scenarios (more restrictive constraints) but, at the same time, globally solvable with standard numerical methods, e.g., CVX \cite{CVX20}. The procedure is iterated until convergence to a local optimum. 

All steps are summarized in Algorithm~\ref{alg:1}, where $\mathbf{d} \equiv \mathbf{d}^{(i)} =[\mathbf{d}_1^{\rm T}, \ldots, \mathbf{d}_K^{\rm T}]^{\rm T}$ denotes the current set of all coefficients, while $\mathbf{d}^{(i-1)}$ refers to the previous solution. Recall that $\mathcal{C}$ was introduced to denote the set of constraints $C1 - C4$.

\subsection{Uplink Power Control}\label{sec:4.2}
Following similar steps, the power control during the UL transmission becomes:
\begin{equation}
\begin{aligned}
    \max_{\{q_k\}, \delta} \, \, & \delta \\
    \textrm{s.t.} \quad & C1: 0 \leq q_k \leq Q_k, \quad \forall k \\ 
    &C2: b_{k,n} q_k \leq E_{k,n}, \quad \forall k,n \\
    &C3: \displaystyle\frac{\tau_u}{\tau_c} B \log_2 \left(1 + \rho_k\left(\{q_k\}\right) \right) \geq \delta, \quad \forall k.
\end{aligned}
\label{eq:17}    
\end{equation}

Unlike before, the newly added constraint $C3$ is quasiconvex (lower bound on the logarithm of a linear fraction; thus, quasiconcave). Accordingly, since $C1$ and $C2$ are linear, for every $\delta$, this will result in a quasilinear problem whose global optimum can be found via CVX. Once again, the optimal value of $\delta$ can be obtained via the bisection method \cite{Boy04}.

\section{Data-Driven Solution: End-to-End Training}\label{sec:5}
In the following, we will derive data-driven implementations to address the former power control optimizations feasibly. Indeed, these approaches will learn from the past theoretical values, a procedure that can be performed offline while keeping the system complexity low for real-time applications \cite{Shl22}.

\begin{algorithm}[t]
\begin{algorithmic}[1]    
    \State Choose convergence tolerance thresholds $\epsilon_1, \epsilon_2 > 0$
    \State Select feasible bisection bounds $\delta_{\rm min}$ and $\delta_{\rm max}$
    \State Initialize coefficients $\mathbf{d}^{(0)} \in \mathcal{C}$  
    \While{$\delta_{\rm max} - \delta_{\rm min} > \epsilon_1$}
    \State Set $\delta = (\delta_{\rm max} + \delta_{\rm min})/2$ and $i = 1$
    \While{$\| \mathbf{d} - \mathbf{d}^{(i - 1)} \|^2/ \| \mathbf{d} \|^2 > \epsilon_2$}
        \State Solve \eqref{eq:16} with $\mathbf{d}^{(i - 1)}$ to find $\mathbf{d}$ using \cite{CVX20}
		\State Set $i = i + 1$ and $\mathbf{d}^{(i - 1)} = \mathbf{d}$   
    \EndWhile
    \If{problem \eqref{eq:16} is feasible}
        \State Set $\delta_{\rm min} = \delta$ and $\mathbf{d}^{(0)} = \mathbf{d}$        
    \Else
        \State Set $\delta_{\rm max} = \delta$
    \EndIf
    \EndWhile
\end{algorithmic}
\caption{SCO-based DL Power Control}
\label{alg:1}
\end{algorithm}

In this section, we start with the most common end-to-end training, where the final solutions $\{p_{k,m}^{\star}\}$ (DL) and $\{q_k^{\star}\}$ (UL) of the model-based approach are considered as the labeled outputs $\mathcal{Y}$ for the DNN \cite{Eli22, Shi23}. The input features $\mathcal{X}$ also differ depending on the scenario, which is why we dedicate separate subsections for the DL and UL architectures. 

\begin{figure*}[t]
\centerline{\includegraphics[scale=1]{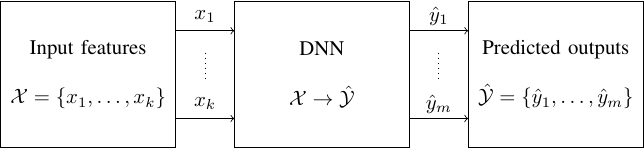}}
\caption{General end-to-end training scheme. The inputs $\mathcal{X}$ are directly mapped onto the outputs $\hat{\mathcal{Y}}$ based on the labels $\mathcal{Y}$.}
\label{fig:2}
\end{figure*}

A general overview of this strategy is provided in Fig.~\ref{fig:2}, where $\hat{\mathcal{Y}}$ is the predicted output, with size $m \triangleq \vert \hat{\mathcal{Y}} \vert$, and $k \triangleq \vert \mathcal{X} \vert$ corresponds to the number of inputs. Note that the features are directly mapped onto the power control coefficients, i.e., a single DNN substitutes the entire optimization.

\subsection{Downlink Architecture}\label{sec:5.1}
Motivated by \cite{Zah23}, instead of using the “raw” channel gains or LSF coefficients as inputs $\mathcal{X}$, we pre-process these features before feeding them to the network. That is, we transform these variables into an initial estimate for $\{p_{k,m}\}$ to facilitate the learning process. In particular, we consider the coefficients derived from the fractional power control (FPC) \cite{Dem21}:
\begin{equation}
     p_{k,m}^{\rm FPC} = P_m \frac{\displaystyle \alpha_{k,m}^{\kappa}}{\displaystyle \sum_{j = 1}^K a_{j,m} \alpha_{j,m}^{\kappa}},
    \label{eq:18}
\end{equation}
where $\alpha_{k,m}$ is the LSF coefficient between AP $m$ and UE $k$, and $\kappa \in [-1,1]$ is the parameter to tune the power distribution: 
\begin{itemize}
    \item $\kappa > 0$ prioritizes good channel qualities, also known as proportional power control,
    \item $\kappa = 0$ provides equal service in the network, i.e., uniform power control (UPC),
    \item $\kappa < 0$ resembles a fairer policy that helps links with poor propagation.
\end{itemize}

This way, the number of inputs and outputs is the same: $\vert \mathcal{X} \vert = \vert \hat{\mathcal{Y}} \vert = N K$ coefficients (recall that $N = \sum_{m = 1}^M a_{k,m}$ is the number of APs serving UE $k$, which is why we have $NK$ elements rather than $MK$). Besides, in line with the discussion in \cite[(7.47)]{Dem21}, we fix $\kappa$ to $0.5$ because opportunistic values are usually preferred in the DL. 

The complete layout of the DNN is reported in Table~\ref{tab:1}. We follow a pyramidal structure, in which we first expand and then compress the features. Unless otherwise stated, all layers are fully connected. Notably, only the first hidden layer has a linear activation function, whereas the rest apply ReLU \cite{Goo16}. This choice has been shown to yield the best performance. 

\begin{table}[t]
\caption{Layout of the DNN in the DL (end-to-end training). Parameters to be trained: 741864.}
\begin{center}
\begin{tabular}{|c|c|c|c|}
\hline
Layer & Size & Parameters & Activation \\ 
\hline
Input & NK & - & - \\
Hidden 1 & 1024 & 41984 & linear \\
Hidden 2 & 512 & 524800 & ReLU \\
Hidden 3 & 256 & 131328 & ReLU \\
Hidden 4 & 128 & 32896 & ReLU \\
Hidden 5 & 64 & 8256 & ReLU \\
Output & NK & 2600 & ReLU \\
\hline
\end{tabular}
\label{tab:1}
\end{center}
\end{table}

To avoid any issues due to bad scaling (samples having a large dynamic range), both features are normalized as follows:
\begin{itemize}
    \item Inputs are normalized according to the interquartile range (IQR), i.e., data is scaled between the first and third quartiles. To further improve the robustness against outliers, we also center the values so that the median is zero.  
    \item Outputs are converted into the logarithmic scale (dB) and scaled through the min-max normalization. This reduces the dynamic range and ensures positive outcomes.
\end{itemize}

Regarding the actual training, the goal is to minimize the mean absolute error (MAE) (also known as $l$-1 loss) between the model-based solution $\mathcal{Y} = \{p_{k,m}^{\star}\}$ from Subsection~\ref{sec:4.1}, and the output predicted by the DNN $\hat{\mathcal{Y}} = \{\hat{p}_{k,m}\}$:
\begin{equation}
    \mathcal{L}_{\rm DL}(\hat{\mathcal{Y}},\mathcal{Y}) = \frac{1}{NK} \sum_{k=1}^K\sum_{m = 1}^M a_{k,m} \vert \hat{p}_{k,m} - p_{k,m}^{\star} \vert,
    \label{eq:19}
\end{equation}
which also helps in the presence of outliers. The loss above is averaged over a total of \num{e5} samples, divided into $80\%$, $10\%$, and $10\%$ for training, validation, and test, respectively.

Moreover, we employ $l_2$ regularization and the Adam optimizer. The learning rate is initially set to \num{e-3} for the first epochs to ensure convergence and gradually decreased by $0.1$ to fine-tune the weights. The maximum epochs are $50$ and the batch size is $256$. Such options are found by trial and error.

\subsection{Uplink Architecture}\label{sec:5.2}
Following the previous rationale, in the UL, we also employ as $K$ input features $\mathcal{X}$ the equivalent FPC coefficients: 
\begin{equation}
    q_k^{\rm FPC} = Q_k \frac{\displaystyle \left(\sum_{m = 1}^M a_{k,m} \alpha_{k,m} \right)^{\varkappa}}{\displaystyle \textrm{max}_j \left(\sum_{m = 1}^M a_{j,m} \alpha_{j,m} \right)^{\varkappa}},
    \label{eq:20}
\end{equation}
where $\varkappa$ is defined as $\kappa$. Conversely, we set $\varkappa = -0.5$ as it truly mirrors a maximin operation \cite[(7.34)]{Dem21}.

The corresponding layout of the DNN is specified in Table~\ref{tab:2}. Similar to before, we follow a pyramidal structure with linear and ReLU activation layers. The main difference is the size of the DNN, provided the number of coefficients equals the number of UEs (now three hidden layers suffice for good accuracy). Accordingly, the MAE is given by:
\begin{equation}
    \mathcal{L}_{\rm UL}(\hat{\mathcal{Y}},\mathcal{Y}) = \frac{1}{K} \sum_{k=1}^K \vert \hat{q}_k - q_k^{\star} \vert,
    \label{eq:21}
\end{equation}
with $q_k^{\star}$ and $\hat{q}_k$ the optimal and learned solutions, respectively. Additionally, we apply the same normalization to the features, and the other training parameters remain unaltered. Only the batch size is decreased to $64$ due to the dimension reduction. 

\begin{table}[t]
\caption{Layout of the DNN in the UL (end-to-end training). Parameters to be trained: 3320.}
\begin{center}
\begin{tabular}{|c|c|c|c|}
\hline
Layer & Size & Parameters & Activation \\ 
\hline
Input & K & - & - \\
Hidden 1 & 64 & 576 & linear \\
Hidden 2 & 32 & 2080 & ReLU \\
Hidden 3 & 16 & 528 & ReLU \\
Output & K & 136 & ReLU \\
\hline
\end{tabular}
\label{tab:2}
\end{center}
\end{table}

\begin{figure*}[t]
\centerline{\includegraphics[scale=1]{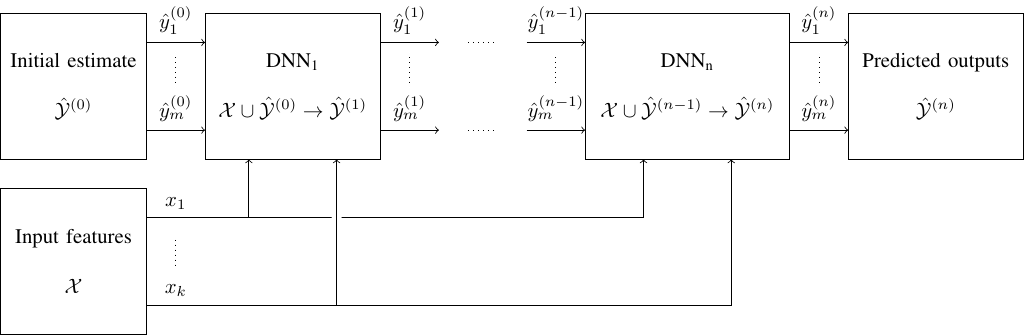}}
\caption{General deep unfolding scheme. The inputs $\mathcal{X}$ are mapped onto the outputs $\hat{\mathcal{Y}} \equiv \hat{\mathcal{Y}}^{(n)}$ via $n$ DNNs (each modeling an SCO iteration).}
\label{fig:3}
\end{figure*}

\section{Data-Driven Solution: Deep Unfolding}\label{sec:6}
Real-time applications typically require low-complexity and flexible solutions \cite{Eld22}. However, in the previous architectures, the DNNs are often treated as “black boxes” \cite{Shl22}. This results in a lack of interpretability that, along with the vast number of parameters and the need for large datasets, questions the online implementation of these end-to-end training strategies.

To circumvent these and other issues, the so-called \textit{deep unfolding} (or algorithm unrolling) emerges as a promising candidate \cite{Mon21}. This approach attempts to mimic the structure of the model-based algorithm by replacing the iterations with small DNNs. In other words, expert knowledge is now incorporated into the design, resulting in higher practical feasibility and further simplification of computational resources. 

Note, however, that this technique is only meaningful in the DL formulation since the optimization in the UL scenario is not sequential, and no unrolling can be applied (cf. \eqref{eq:17}). Therefore, we will address each subproblem in \eqref{eq:16} through separate concatenated DNNs. Accordingly, the $i$-th DNN will take as input the set of channels $\mathcal{X}$ together with the previous power coefficients $\hat{\mathcal{Y}}^{(i - 1)}$. This working principle is shown in Fig.~\ref{fig:3}, with $n$ the total number of iterations.

Unfortunately, the bisection search cannot be directly modeled through deep unfolding, given that feasibility problems are ill-defined in DNN architectures \cite{Pel22}. In the following, we will present an alternative SCO-based method for finding a feasible solution that can be unfolded.

\subsection{Unfolded Downlink Power Control}\label{sec:6.1}
The DL power control can be equivalently rewritten as
\begin{equation}
\begin{aligned}
        \max_{\{d_{k,m}\}} \, \min_k \, \, & \rho_k \\
        \textrm{s.t.} \quad & C1: d_{k,m} \geq 0, \quad \forall k,m\\ 
        &C2: \sum_{k = 1}^K a_{k,m} d_{k,m}^2 \leq P_m, \quad \forall m \\ 
        &C3: \sum_{j = 1}^K \left\vert \sum_{m = 1}^M a_{j,m} d_{j,m} \mathbf{h}_{k,m}^{\rm H} \mathbf{b}_{j,m} \right\vert^2 \leq \frac{\lambda^2 I_k}{4 \pi}, \, \forall k.
    \end{aligned}
    \label{eq:22}      
\end{equation}

In line with \cite[Subsection~VI-B]{Zhi23}, the min function can be approximated by the LSE (log-sum-exp), i.e.,
\begin{equation}
    \textrm{min} \{x_1,\ldots,x_k\} \geq -\frac{1}{\upsilon} \ln\left(  e^{- \upsilon x_1} + \ldots + e^{- \upsilon x_k} \right),
    \label{eq:23}
\end{equation}
where $\upsilon > 0$ controls the accuracy. 

Translated to our problem, we can then safely replace the objective in \eqref{eq:22} with \eqref{eq:23}:
\begin{equation}
    \min_k \, \, \rho_k\left(\mathbf{d}\right) \geq -\frac{1}{\upsilon} \ln\left(  \sum_{k = 1}^K \exp\left(-\upsilon \rho_k\left(\mathbf{d} \right) \right) \right) \triangleq \varrho\left(\mathbf{d}\right).
    \label{eq:24}
\end{equation}

However, although the approximation in \eqref{eq:24} is concave w.r.t. the arguments $\rho_k(\mathbf{d})$ \cite{Boy04}, we are interested in designing the coefficients $\mathbf{d}$. Therefore, since concavity is not preserved in this transformation, we must revert to SCO and derive a concave lower bound for \eqref{eq:24}.

Let us start by analyzing the concavity of $\varrho(\mathbf{d})$, which is indeed equivalent to $-\varrho(\mathbf{d})$ being convex. Without loss of generality, we set $\upsilon = 1$ in the rest of this subsection.

To ensure $-\varrho(\mathbf{d})$ is convex, we need all the addends to be log-convex, as this property is maintained under sums \cite{Boy04}. This is straightforward to verify in the case of $\rho_k(\mathbf{d})$ provided that logarithms and exponentials are inverse functions, i.e., $\ln(\exp(-\rho_k(\mathbf{d}))) = -\rho_k(\mathbf{d})$, $\forall k$ (linear). 

Expanding the actual expression of the SINR defined in \eqref{eq:3}, we have
\begin{equation}
    \begin{aligned}
        -\rho_k(\mathbf{d}) &= \frac{\displaystyle -\left\vert \sum_{m = 1}^M a_{k,m} d_{k,m} \mathbf{h}_{k,m}^{\rm H} \mathbf{b}_{k,m} \right\vert^2}{\displaystyle \sum_{j \neq k}\left\vert \sum_{m = 1}^M a_{j,m} d_{j,m} \mathbf{h}_{k,m}^{\rm H} \mathbf{b}_{j,m} \right\vert^2 + \sigma_k^2},
    \end{aligned}    
    \label{eq:25}
\end{equation}
which is equivalent to the negative ratio of the two convex functions already defined in \eqref{eq:14}, $g_k(\mathbf{d}_k)$ and $f_k(\bar{\mathbf{d}}_k)$. Hence, this means \eqref{eq:25} is certainly not convex w.r.t. $\mathbf{d}$.

Luckily, \eqref{eq:25} has the same structure as the constraint in \eqref{eq:14}: we can thus linearize the numerator in \eqref{eq:25} at the previous feasible point. More precisely, by using the bound in \eqref{eq:15}, we obtain the following surrogate function for the SCO:
\begin{equation}
    -\rho_k(\mathbf{d}) \leq - \frac{\displaystyle\tilde{g}_k\left(\mathbf{d}_k,\mathbf{d}_k^{(i - 1)}\right)}{\displaystyle f_k\left(\bar{\mathbf{d}}_k\right)}.
    \label{eq:26}
\end{equation}

This way, since \eqref{eq:26} is now convex (linear over quadratic is concave), the resulting objective function at the $i$-th iteration will be concave, namely
\begin{equation}    
    \begin{aligned}
    \varrho(\mathbf{d}) &\geq - \ln\left(  \sum_{k = 1}^K \exp\left(- \frac{\displaystyle \tilde{g}_k\left(\mathbf{d}_k,\mathbf{d}_k^{(i - 1)}\right)}{\displaystyle f_k\left(\bar{\mathbf{d}}_k\right)}\right)\right) \\ &\triangleq \tilde{\varrho}\left(\mathbf{d},\mathbf{d}^{(i - 1)}\right).    
    \end{aligned}
    \label{eq:27}
\end{equation}

Under those assumptions, we finally end up with a sequence of convex subproblems that can be globally solved and converge to a local optimum of \eqref{eq:22}, i.e.,
\begin{equation}    
    \max_{\mathbf{d}} \, \, \tilde{\varrho}\left(\mathbf{d},\mathbf{d}^{(i - 1)}\right) \quad \textrm{s.t.} \quad C1-C3.    
    \label{eq:28}
\end{equation}

The procedure is reported in Algorithm~\ref{alg:2}, whose accuracy will be numerically assessed in Subsection~\ref{sec:7.6}.

\begin{algorithm}[t]
\begin{algorithmic}[1]    
    \State Initialize coefficients $\mathbf{d}^{(0)} \in \mathcal{C}$    
    \State Choose convergence tolerance threshold $\epsilon_3$  
    \State Set $i = 1$
    \While{$\| \mathbf{d} - \mathbf{d}^{(i - 1)} \|^2/ \| \mathbf{d} \|^2 > \epsilon_3$}
        \State Solve \eqref{eq:28} with $\mathbf{d}^{(i - 1)}$ to find $\mathbf{d}$ using \cite{CVX20}
		\State Set $i = i + 1$ and $\mathbf{d}^{(i - 1)} = \mathbf{d}$       
    \EndWhile
\end{algorithmic}
\caption{Unfolded SCO-based DL Power Control}
\label{alg:2}
\end{algorithm}

\subsection{Unfolded Downlink Architecture}\label{sec:6.2}
\begin{figure*}[t]
\centerline{\includegraphics[scale=1]{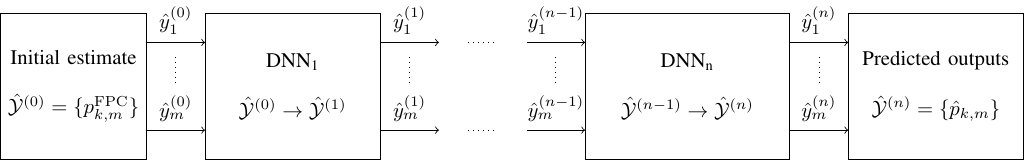}}
\caption{Deep unfolding scheme without side information. The initial estimate $\hat{\mathcal{Y}}^{(0)}$ is sequentially updated until the final output $\hat{\mathcal{Y}}^{(n)}$.}
\label{fig:4}
\end{figure*}

Thanks to the previous formulation, we can now unroll the model-based solution and implement a deep unfolding scheme. Following the reasoning in Section~\ref{sec:5}, we also consider as input features the FPC coefficients from \eqref{eq:18}. This means we can remove the side information in the general scheme depicted in Fig.~\ref{fig:3} and start directly from the estimate $\hat{\mathcal{Y}}^{(0)} = \{p_{k,m}^{\rm FPC}\}$. This particularization is shown in Fig.~\ref{fig:4}.

The initial estimate is used as input for the first DNN, which is trained to minimize the MAE between the prediction $\mathcal{Y}^{(1)} = \{\hat{p}_{k,m}^{(1)}\}$ and the ultimate labels $\mathcal{Y} = \{p_{k,m}^{\star}\}$:
\begin{equation}
    \mathcal{L}_{\rm DL}^{(1)}(\hat{\mathcal{Y}}^{(1)},\mathcal{Y}) = \frac{1}{NK} \sum_{k=1}^K\sum_{m = 1}^M a_{k,m} \vert \hat{p}_{k,m}^{(1)} - p_{k,m}^{\star} \vert.
    \label{eq:29}
\end{equation}

Next, the output $\{\hat{p}_{k,m}^{(1)}\}$ is used as input for the subsequent DNN, and we then repeat the training process to obtain the prediction $\{\hat{p}_{k,m}^{(2)}\}$ that minimizes the loss $\mathcal{L}_{\rm DL}^{(2)}(\hat{\mathcal{Y}}^{(2)},\mathcal{Y})$. This is done until we reach the maximum number of iterations $n$. 

Thanks to this sequential learning, the DNNs can be composed of fewer layers. All follow a pyramidal structure similar to those presented before, and the same feature normalization is applied. The complete layout is given in Table~\ref{tab:3}.

\begin{table}[t]
\caption{Layout of the $i$-th DNN in the DL (deep unfolding). Parameters to be trained (per iteration): 10408.}
\begin{center}
\begin{tabular}{|c|c|c|c|}
\hline
Layer & Size & Parameters & Activation \\ 
\hline
Input & NK & - & - \\
Hidden 1 & 128 & 5248 & linear \\
Output & NK & 5160 & ReLU \\
\hline
\end{tabular}
\label{tab:3}
\end{center}
\end{table}

As a result, the benefits of deep unfolding are twofold: (1) we can adapt the system architecture to specific problems with domain knowledge; and (2) we can easily control the algorithm complexity by adjusting the number of iterations. As we will see later in the simulations, $n \leq 3$ iterations normally suffice for good performance so that the total number of parameters to be trained is significantly less than those needed with the end-to-end approach described in Subsection~\ref{sec:5.1}.

\section{Numerical Simulations}\label{sec:7}
Numerical simulations will assess the communication performance and computational complexity of the previous power control schemes. Thanks to that, we will be able to evaluate the pros and cons of the model-based and data-driven solutions. In that sense, we start by assessing the efficacy of the end-to-end training for both DL and UL described in Section~\ref{sec:5}. After that, we will shift the focus toward the deep unfolding approach (for the DL only). Accordingly, we will also validate the accuracy of the LSE approximation derived in Section~\ref{sec:6}.

Before proceeding, we dedicate some initial subsections to presenting the channel model, channel estimation, receive and transmit signal processing, and system parameters.

Last, note that the upcoming insights are consistent with the outcomes from our preliminary work \cite{Lie24}, where an equivalent multi-cell mMIMO was also analyzed. Cell-free was shown to surpass cellular deployments, which is why in this paper, we concentrate on the former. As a matter of fact, CF-mMIMO poses more design challenges given that the number of (DL) power coefficients is significantly larger. Hence, the distributed architecture can be seen as a more general framework.

\subsection{Propagation Channel}\label{sec:7.1}
The link between UE $m$ and AP $k$ is \cite[(1)]{Elw23}
\begin{equation}
    \mathbf{h}_{k,m} = \sqrt{\frac{\alpha_{k,m}}{1 + \beta_{k,m} }}\left(\sqrt{\beta_{k,m}}e^{j \psi_{k,m}} \mathbf{v}_m\left(\theta_{k,m}\right) + \bar{\mathbf{h}}_{k,m} \right),
    \label{eq:30} 
\end{equation}
where $\alpha_{k,m}$ is the LSF coefficient including the path loss, $\beta_{k,m}$ is the Rician factor, $[\bar{\mathbf{h}}_{k,m}]_l \sim \mathcal{CN}(0,1)$ are the uncorrelated Rayleigh-distributed non-line-of-sight (NLoS) components, $\psi_{k,m} \sim \mathcal{U}[0,2\pi]$ is the phase offset, $\mathbf{v}_m(\cdot) \in \mathbb{C}^{L}$ is the steering vector (generated according to a uniform linear array), and $\theta_{k,m}$ is the corresponding (LoS) angle of arrival.

Following the discussion in \cite{DAn20}, the Rician factors depend on the probability of LoS of each link, i.e.,
\begin{equation}
    \beta_{k,m} = \frac{p_{\rm LoS}\left(\zeta_{k,m}\right)}{1 - p_{\rm LoS}\left(\zeta_{k,m}\right)},
    \label{eq:31}
\end{equation}
where $p_{\rm LoS}(\zeta_{k,m})$ is a function of the distance $\zeta_{k,m}$ from UE $k$ to AP $m$ \cite[Table B.1.2.1-2]{3GPP36814}.

\subsection{Linear Uplink Channel Estimation}\label{sec:7.2}
Perfect CSI might be an unrealistic assumption in practical systems. Instead, we must acquire this knowledge locally at the APs via UL orthogonal pilots. This allows us to characterize the sufficient statistics of the channels \cite{Ngo17}.

After some manipulations, the following linear minimum mean-squared error estimates can be constructed:
\begin{equation}
    \hat{\mathbf{h}}_{k,m} = \sqrt{\tau_p \mu_k}\mathbf{C}_{k,m} \mathbf{D}_{k,m}^{-1} \mathbf{u}_{k,m},
    \label{eq:32}
\end{equation}
where $\mu_k$ is the training pilot's power,
\begin{equation}
    \mathbf{C}_{k,m} = \frac{\alpha_{k,m}}{1 + \beta_{k,m} } \left(\beta_{k,m} \mathbf{v}_m\left(\theta_{k,m}\right)\mathbf{v}_m^{\rm H}\left(\theta_{k,m}\right) + \mathbf{I}_L \right),
    \label{eq:33}
\end{equation}
refers to the covariance matrix of $\mathbf{h}_{k,m}$, and
\begin{equation}
    \mathbf{D}_{k,m} = \sum_{j = 1}^K \tau_p \mu_j \mathbf{C}_{j,m} \left\vert \mathbf{t}_j^{\rm H} \mathbf{t}_k \right\vert^2 + \eta_m^2 \mathbf{I}_L,
    \label{eq:34}
\end{equation}
is the covariance matrix of the observation $\mathbf{u}_{k,m} \in \mathbb{C}^L$, with $\mathbf{t}_k \in \mathbb{C}^{\tau_p}$ the training sequence of length $\tau_p$ sent by UE $k$. For more details, please refer to \cite[Subsection~II-C]{DAn20}.

This way, we can incorporate CSI errors into the QoS power control. In short, we will replace the channels in the SINRs \eqref{eq:3} and \eqref{eq:6} by their estimates and investigate the performance under conjugate beamforming (CB) \cite{Dem21}, i.e.,
\begin{equation}
    \mathbf{b}_{k,m} = \hat{\mathbf{h}}_{k,m}/\| \hat{\mathbf{h}}_{k,m} \|, \quad \mathbf{f}_{k,m} = \hat{\mathbf{h}}_{k,m},
    \label{eq:35}
\end{equation}
which entails the lowest complexity among the standard linear processing schemes. However, more complex techniques can also be chosen, as the structure of our solutions is not altered. As outlined later, in some UL settings, we will also consider regularized zero forcing (RZF) \cite{Zah23}:
\begin{equation}
    \mathbf{f}_{k,m} = \mu_k \left(\sum_{j = 1}^K \mu_j \hat{\mathbf{h}}_{j,m}\hat{\mathbf{h}}_{j,m}^{\rm H} + \sigma_k^2 \mathbf{I}_L \right)^{-1}\hat{\mathbf{h}}_{k,m}.
    \label{eq:36}
\end{equation}

For simplicity, since the goal of this paper is to compare the performance of model-based and data-driven power controls, we will generally employ CB for precoders/filters.

\subsection{System Parameters}\label{sec:7.3}
Throughout all experiments, we consider a deployment area of $0.5$ km\textsuperscript{2}, wrapped around the edges to avoid boundary effects. The setting follows the micro-urban configuration described in \cite{3GPP36814} with $P_m = 23$ dBm, $Q_k = \mu_k = 20$ dBm, $\sigma_k^2 = \eta_m^2 = N_o B$, $N_o = -174$ dBm/Hz, and $B = 20$ MHz. Unless otherwise stated, $K = 8$ UEs and $M = 16$ APs with $L = 4$ antennas are randomly located within the scenario at fixed heights of $1.65$ m and $10$ m, respectively. Besides, we set $N = 5$ for the UE-AP association.

In line with the findings in \cite{Chi21}, we concentrate on whole-body EMF constraints applied to the general public: $I_k = 10$ W/m\textsuperscript{2} for the IPD limit and $E_{k,n} = 0.08$ W/kg for the SAR (single) metric, with coefficient $b_{k,n} = 8$ kg\textsuperscript{-1}.

Finally, we assume block fading with OFDM modulation and, unlike other works (e.g., \cite{Zah23}), the presence of shadowing. This means our proposal is robust to slow fading variations, whose spatial inconsistency can be critical for DNNs' training. The coherence time and bandwidth are $1$ ms and $200$ kHz, respectively; $\tau_c = 200$ time-frequency samples are thus available for communication \cite{Elw23}. Accordingly, the first $\tau_p = K/2$ symbols will be dedicated to UL channel estimation (i.e., pilot contamination is included), and $\tau_d = \tau_u = (\tau_c - \tau_p)/2$ samples will be assigned to DL and UL transmissions (cf. \eqref{eq:9} and \eqref{eq:11}). 

\subsection{Downlink Performance}\label{sec:7.4}
Along with the optimal power control (OPC) from Subsection~\ref{sec:4.1} and the end-to-end (E2E) DNN-based algorithm from Section~\ref{sec:5}, we also include the UPC plus the FPC given in \eqref{eq:18} with $\kappa = -0.5$ (fair) and $\kappa = 0.5$ (opportunistic) as benchmark schemes. This will help to emphasize the performance of our proposal.

\begin{figure}[t]            
    \centerline{\includegraphics[scale=1]{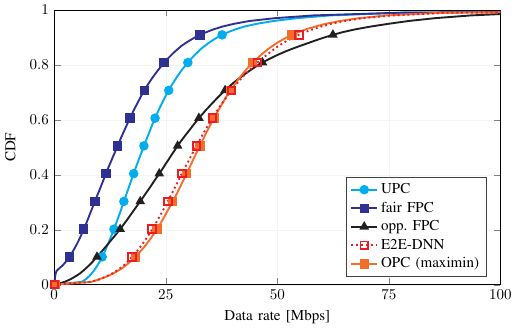}}
    \caption{CDF of the DL user rate under CB processing with UPC, fair FPC, opportunistic (opp). FPC, OPC (maximin), and E2E-DNN.}
    \label{fig:5}
\end{figure}

The CDF (cumulative distribution function) of the UE data rate is depicted in Fig.~\ref{fig:5}. As expected, the OPC outperforms the other policies, especially in terms of unlucky UEs. In addition, the DNN-based technique achieves similar results. For instance, both power control mechanisms ensure better QoS for around $70\%$ of users compared to that of the opportunistic FPC. This motivates the use of maximin optimizations in the DL to generate a database for E2E training (only a small deviation between both curves is obtained).

\begin{figure}[t]            
    \centerline{\includegraphics[scale=1]{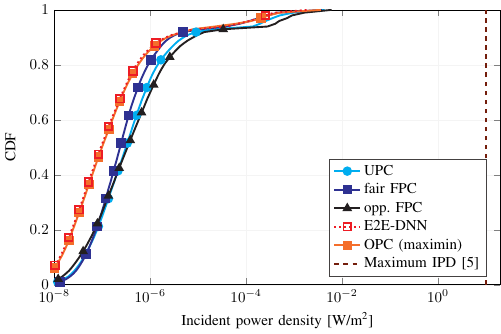}}
    \caption{CDF of the IPD user radiation under CB processing with UPC, fair FPC, opp. FPC, OPC (maximin), and E2E-DNN.} 
    \label{fig:6}    
\end{figure}

The CDF of the radiation per UE (measured in terms of IPD) is presented in Fig.~\ref{fig:6}. For this particular setting, the EMF constraint does not seem to impact the QoS (the maximum IPD limit is far beyond the perceived exposure). This phenomenon is due to the high propagation losses in the DL, which again emphasizes the safety of the CF-mMIMO architecture (cf. \cite{Lie24}). Once more, the OPC yields the smallest IPDs, and the E2E-DNN provides almost the same values.

Overall, at the expense of a slight throughput deterioration, one can state that data-driven methods can properly substitute model-based approaches. This is more relevant in real-time applications, since the learning process is executed offline and the DNN prediction is practically immediate. Further details on the complexity analysis are given in Subsection~\ref{sec:7.7}.

\subsection{Uplink Perfomance}\label{sec:7.5}
\begin{figure}[t]            
    \centerline{\includegraphics[scale=1]{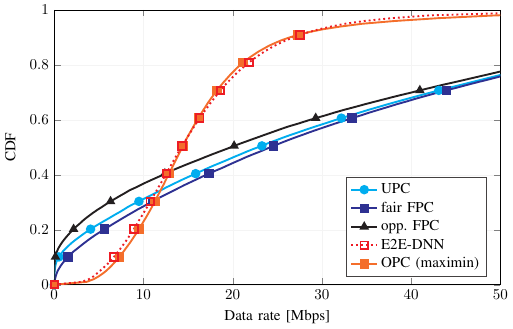}}
    \caption{CDF of the UL user rate under CB processing with UPC, fair FPC, opp. FPC, OPC (maximin), and E2E-DNN.}
    \label{fig:7}    
\end{figure}

\begin{figure}[t]        
    \centerline{\includegraphics[scale=1]{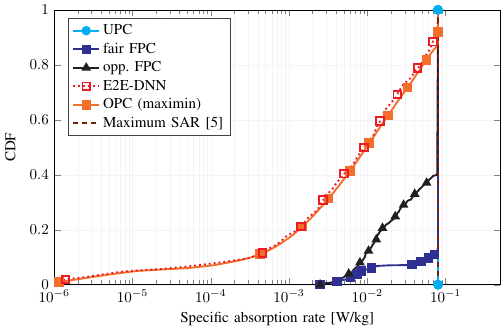}}
    \caption{CDF of the SAR user radiation under CB processing with UPC, fair FPC, opp. FPC, OPC (maximin), and E2E-DNN.} 
    \label{fig:8}    
\end{figure}

The CDFs of the UL data rate and the UE perceived SAR are depicted in Figs.~\ref{fig:7} and \ref{fig:8}, respectively. Once again, the OPC solution yields higher rates than its heuristic counterparts, and the E2E-trained DNN achieves close QoS. Similarly, both strategies outperform the other power control mechanisms in terms of EMF exposure. However, unlike before, the SAR limit is no longer negligible, especially for the UPC and FPC. It is noteworthy that both policies are designed to comply with the EMF restriction: we scale the coefficients in \eqref{eq:20} so that the SAR constraint is fulfilled. For instance, in the UPC, the ultimate transmit power will indeed be $\max (Q_k, E_{k,n}/b_{k,n})$. That is to say, an unconstrained allocation could easily exceed the maximum radiation established by ICNIRP and FCC. This underscores the interest in EMF analysis for the UL. 

\begin{figure}[t]    
    \centerline{\includegraphics[scale=1]{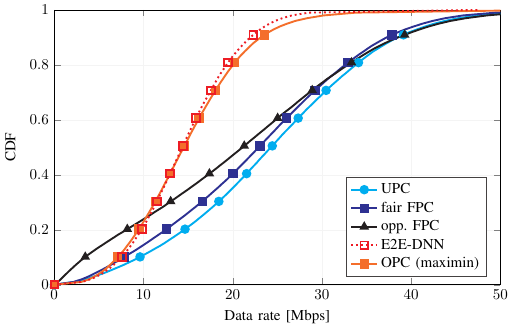}}
    \caption{CDF of the UL user rate under RZF processing with UPC, fair FPC, opp. FPC, OPC (maximin), and E2E-DNN.}
    \label{fig:9}
    \end{figure}

\begin{figure}[t]    
    \centerline{\includegraphics[scale=1]{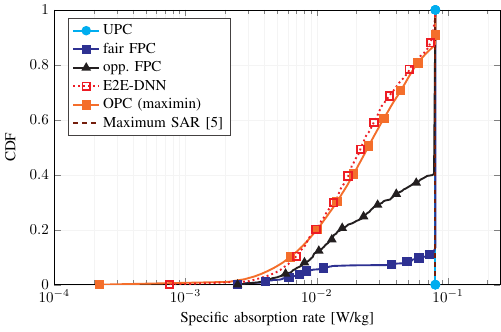}}
    \caption{CDF of the SAR user radiation under RZF processing with UPC, fair FPC, opp. FPC, OPC (maximin), and E2E-DNN.} 
    \label{fig:10}    
\end{figure}

The results under RZF processing are pictured in Figs.~\ref{fig:9} and \ref{fig:10}. The main difference w.r.t. the CB setting lies in the performance of the heuristic policies. Since interference is mitigated, UPC and FPC mechanisms can attain a good QoS while maintaining the same EMF radiation (their coefficients are independent of the spatial filters). Contrarily, the benefit for the OPC is small, yet the E2E-DNN still performs well; thus, our proposal can adapt to different beamforming schemes.


\begin{figure}[t]    
    \centerline{\includegraphics[scale=1]{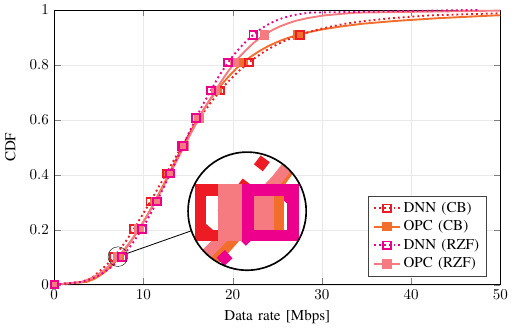}}
    \caption{CDF of the UL user rate under CB and RZF processing with OPC (maximin) and E2E-DNN.}
    \label{fig:11}
\end{figure}

For clarity in the explanation, the throughput of the model-based and data-driven methods is explicitly compared in Fig.~\ref{fig:11}. As mentioned, the impact of the receive combiner is only noticeable in the upper part of the CDF, corresponding to the UEs with strong links. Thanks to the interference reduction of RZF processing, UEs can transmit with more power and obtain a slightly more uniform service over the deployment area \cite{Dem21}. However, as discussed below, the increase in power also yields an increase in EMF exposure. 

\begin{figure}[t]        
    \centerline{\includegraphics[scale=1]{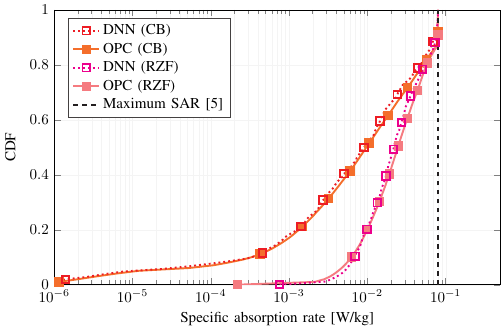}}
    \caption{CDF of the SAR user radiation under CB and RZF processing with OPC (maximin) and E2E-DNN.} 
    \label{fig:12}
\end{figure}

To further investigate the role of beamforming, the radiation of the OPC and E2E-DNN is also reported in Fig.~\ref{fig:12}. We can observe that, while being below the SAR limit, RZF entails a higher EMF exposure due to the power increase. This indicates that, under maximin operation, CB filters are more suited for handling the UL setup: we readily satisfy the health regulations without substantially compromising the data rate. However, when resorting to heuristic policies, RZF clearly outperforms CB since powers and, thus, SARs, remain unaltered. 

Lastly, recall that RZF is only illustrated in the UL setting. The reason is that, as stated in the previous subsection, the EMF constraint does not deteriorate the throughput in the DL (the IPD limit is far away from the user radiation). Hence, no relevant insights can be extracted: the RZF scheme always surpasses the communication performance of CB (cf. \cite{Dem21}).

\subsection{Unfolded Downlink Perfomance}\label{sec:7.6}
In Figs.~\ref{fig:13} and \ref{fig:14}, we portray the CDFs of the user data rate and EMF exposure corresponding to the unfolded DL power control. More precisely, we include the LSE-based OPC and the unfolded DNN (U-DNN) together with the maximin OPC and the conventional E2E-DNN. Recall that ($n$) denotes the number of iterations in the U-DNN algorithm.

At a glance, one can assert that the LSE approximation given in \eqref{eq:24} is tight, i.e., the QoS performance is practically equal in both OPC solutions. The same holds in the case of the perceived IPD: a similar radiation is obtained. 

On the other hand, the throughput is enhanced with the number of iterations or U-DNNs. This is not surprising because we are gradually updating the power coefficients until convergence to the stationary point (described by the OPC). In that sense, the EMF radiation also decreases progressively with the number of steps (as indicated by the dashed lines). 

In summary, these results validate the accuracy of the LSE-based power control and the efficacy of the deep unfolded technique. That is, incorporating expert knowledge into the training strategy contributes to the learning process while maintaining a low level of complexity. This suggests that algorithm unrolling is a potential candidate for real-time applications.

\subsection{Complexity Analysis}\label{sec:7.7}
One of the main advantages of convex problems is that they can be solved in polynomial time \cite{Boy04}. In terms of big-O, this means that the computational cost scales polynomially with the number of design variables. Accordingly, the complexity per iteration of the bisection search in the SCO-based DL power control (i.e., line 7 in Algorithm~\ref{alg:1}) is $\mathcal{O}(NK)$. 

\begin{figure}[t]        
    \centerline{\includegraphics[scale=1]{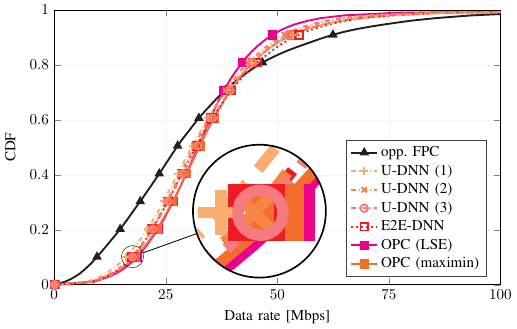}}
    \caption{CDF of the DL user rate under CB processing with opp. FPC, OPC (maximin), OPC (LSE), E2E-DNN, and U-DNN.}
    \label{fig:13}
\end{figure}

Similarly, the cost of each iteration in the unfolded method (i.e., line 5 in Algorithm~\ref{alg:2}) is again $\mathcal{O}(NK)$. However, since no bisection is employed, the number of required iterations is possibly lower, and so is the overall execution time. This can be verified in Table~\ref{tab:1}, where we present the average execution time of each solution. Hence, one can state that the LSE-based approach surpasses the traditional SCO technique for maximin policies: the total running time can be significantly decreased (around three times) at the expense of a negligible performance loss in terms of data rate (cf. Fig.~\ref{fig:13}).

In the case of the UL setting, the computational complexity per iteration is smaller, i.e., $\mathcal{O}(K)$. Unsurprisingly, this yields a faster execution, as shown in Table~\ref{tab:4}. Note that the average time needed by both processing schemes (CB and RZF) is almost the same, as the iteration cost scales equally.

On the other hand, once the learning tasks are completed offline (that is, generating the database by solving the previous optimizations and training the neural networks), the running time of the data-driven methods depends on the specific layout. In particular, for a DNN with $n$ fully-connected layers, each with $m_i$ neurons, the required number of real multiplications and additions at the $i$-th layer is $m_i$ and $m_{i - 1}$, respectively, with $i\in\{1,\ldots,n\}$. \cite[Subsection VI~C]{Zah23}. Apart from that, a total of $\sum_{i=1}^n m_i$ activation functions must also be evaluated.

As illustrated in Table~\ref{tab:4}, all learning-based routines entail a far lower running time. More precisely, their execution lies in the order of milliseconds (more than $1000$ times faster than that of the classical solutions). Consequently, since these data-driven alternatives are shown to provide a suitable throughput and radiation performance, they can effectively replace their theoretical counterpart, especially in real-time applications. 

\begin{figure}[t]            
    \centerline{\includegraphics[scale=1]{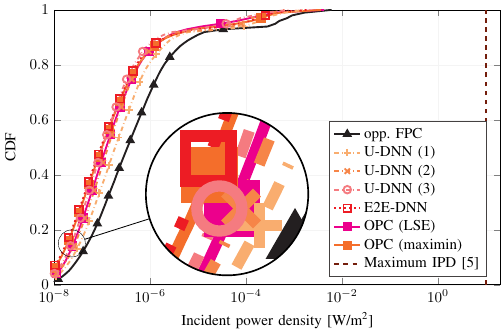}}
    \caption{CDF of the IPD user radiation under CB processing with opp. FPC, OPC (maximin), OPC (LSE), E2E-DNN, and U-DNN.} 
    \label{fig:14}    
\end{figure}

Despite that, there is a noticeable gap between the different approaches. Remarkably, in the UL, CB processing is a bit slower than ZF filtering because the solution in interference-limited scenarios might be less trivial. In other words, the learning process with interference-cancellation techniques becomes easier, which is certainly a meaningful and worthwhile advantage. When looking at the DL setting, one can clearly see that deep unfolding outperforms conventional E2E-DNNs. For instance, the unrolled architecture with just one iteration, denoted by U-DNN(1), needs less than five times the execution time of E2E-DNN. Recall that this choice already behaves well in terms of data rate and EMF exposure, as discussed in Figs.~\ref{fig:13} and \ref{fig:14}, respectively. Therefore, when it comes to the learning of sequential procedures like SCO, there is still room for efficiency improvement with deep unfolding.

\begin{table*}[t]
\caption{Average execution time of the model-based and data-driven power control algorithms (in seconds). Programs written in MATLAB R2024a and implemented on a Windows 10 x64 machine with 64 GB RAM and an AMD Ryzen 9 5950X 16-Core CPU running at 4 GHz.}
\begin{center}
\begin{tabular}{|c|c|c|c|c|c|c|}
\hline
Setting & OPC (maximin) & OPC (LSE) & E2E-DNN & U-DNN (1) & U-DNN (2) & U-DNN (3) \\ 
\hline
DL (CB) & \num{60.04} & \num{19.07} & \num{10.26e-3} & \num{1.89e-3} & \num{4.54e-3} & \num{6.90e-3} \\
UL (CB) & \num{4.46} & - & \num{2.17e-3} & - & - & - \\
UL (RZF) & \num{4.39} & - & \num{1.68e-3} & - & - & - \\
\hline
\end{tabular}
\label{tab:4}
\end{center}
\end{table*}

\section{Conclusions}\label{sec:8}
This paper has addressed power control in user-centric CF-mMIMO systems with EMF exposure constraints, using both model-based and data-driven approaches. The model-based solutions for the DL have relied on SCO techniques and the LSE (log-sum-exp) approximation, while the UL problem has been tackled using standard convex optimization tools. In parallel, data-driven strategies have been investigated, including both E2E (end-to-end) deep learning architectures and deep unfolding methods. The results have shown that the proposed model-based methods effectively enforce the EMF constraints while delivering excellent performance in terms of user fairness. Moreover, the data-driven approaches have been demonstrated to closely approximate the performance of the model-based solutions, with the added benefit of significantly lower computational complexity.

Overall, our findings have confirmed the viability of integrating EMF-aware power control into CF-mMIMO systems and have highlighted the potential of learning-based methods to provide low-complexity and efficient solutions in practical deployments.

\balance
\bibliographystyle{IEEEtran}
\bibliography{references}

\end{document}